\documentclass{article}
\usepackage{spconf,amsmath,graphicx}
\usepackage{algorithmic}
\usepackage{booktabs,subfigure,float,amsmath}
\usepackage[ruled,linesnumbered]{algorithm2e}
\usepackage{color}

\title{AutoLV: Automatic lecture video generator}
\name{Wenbin Wang \quad Yang Song \quad Sanjay Jha}

\address{School of Computer Science and Engineering, University of New South Wales, Australia}
\begin{document}
%

\maketitle
\begin{abstract}
We propose an end-to-end lecture video generation system that can generate realistic and complete lecture videos directly from annotated slides, instructor’s reference voice and instructor’s reference portrait video. Our system is primarily composed of a speech synthesis module with few-shot speaker adaptation and an adversarial learning-based talking-head generation module. It is capable of not only reducing instructors’ workload but also changing the language and accent which can help the students follow the lecture more easily and enable a wider dissemination of lecture contents. Our experimental results show that the proposed model outperforms other current approaches in terms of authenticity, naturalness and accuracy. Here is a video demonstration of how our system works, and the outcomes of the evaluation and comparison: https://youtu.be/cY6TYkI0cog.
\end{abstract}
\begin{keywords}
speech synthesis, talking-head generation, e-learning. 
\end{keywords}

\vspace{-0.5cm}
\section{Introduction}
\vspace{-0.3cm}
\label{sec:intro}

Distance education, particularly Massive Open Online Courses (MOOCs), has transformed the way people throughout the globe learn knowledge in recent years \cite{calvo2020educating}. Especially given the current COVID-19 situation, online video lectures provide a much safer alternative to face-to-face lectures. 
Subsequently, methods for enhancing students' experience of watching lecture videos via artificial intelligence technologies have been proposed, e.g., repacking the text, graphics, charts and instructor's speech voice in the lecture video into an interactive notebook structure via multimedia analysis to enhance students' 
learning efficiency \cite{DBLP:conf/icmcs/XuWLLZSH19}, and automatically retrieving lecture videos to assist students in understanding knowledge points via real-time speech recognition \cite{matsumoto2019ai}. However, existing methods have been designed to improve students' experience, whereas little work has been done to 
improve instructors' experience of recording and updating lecture videos. Although the substance of lectures may not vary much from semester to semester, in the majority of disciplines, instructors may spend significant time re-recording and editing videos each semester, which imposes a significant burden on their workload. 

A complete lecture video typically contains three components: presentation slides, instructor's talking-head video and audio recording of the lecture. Although the instructor should carefully design the slides through their professional knowledge, we can free them from recording the speech audio and talking-head video, so that they have more time to devote to other work. In our application, given the slides as input, we endeavor to synthesize talking-head video and audio automatically from the slides. 
However, achieving this goal needs to address many challenges. For example, a limited amount of voice training data would reduce the quality and naturalness of the synthesized voice; incorrect synthesis of less-known professional terminologies may lead to wrong pronunciation; and using the basic video splicing approach would result in incoherent head movement in long-duration talking-head videos.

To tackle these challenges, we combine speech synthesis with few-shot speaker adaptation and a generative adversarial network (GAN) model for talking-head generation with video augmentation to automate our entire system. 
Our system can generate realistic and complete lecture videos directly from annotated slides using only a few minutes of the instructor's voice recording and seconds of the instructor's portrait video. Our method also enables efficient translation of lectures into other languages using the neural machine translation module  \cite{DBLP:conf/nips/VaswaniSPUJGKP17}.
Furthermore, students could personalize lecture videos themselves to adapt to their learning styles. For example, students who are not familiar with the British accent could personalize lecture videos with their familiar accents.

This paper presents the following contributions:
\vspace{-0.2cm}
\begin{itemize}
\item[$\bullet$] We propose an end-to-end deep learning method that helps instructors generate lecture videos automatically and allows students to customize lecture videos. The system significantly reduces the instructors' workload and helps fulfill students' diverse preferences for lecture videos.
\vspace{-0.3cm}
\end{itemize}

\begin{itemize}
\item[$\bullet$] A few-shot speaker adaptation method is applied on top of an off-the-shelf speech synthesizer to achieve high-quality voice cloning with only a few minutes of reference voice data. It can generate highly accurate and natural synthesis of speech.
\vspace{-0.3cm}
\end{itemize}

\begin{itemize}
\item[$\bullet$] A video temporal augmentation algorithm is designed to solve the problem of incoherent head movement caused by insufficient reference video when generating a long talking-head video. It also contributes to the naturalness of the generated talking-head video.
\vspace{-0.3cm}
\end{itemize}

\vspace{-0.4cm}
\section{RELATED WORK}
\vspace{-0.2cm}
\label{sec:pagestyle}

\subsection{Neural Speech Synthesis}
\vspace{-0.2cm}
With the rapid development of deep learning technologies, the performance of text-to-speech method has improved significantly. Tacotron \cite{DBLP:conf/interspeech/WangSSWWJYXCBLA17} and its enhanced methods \cite{DBLP:conf/icassp/ShenPWSJYCZWRSA18, DBLP:conf/icml/WangSZRBSXJRS18, DBLP:conf/interspeech/YuLH0WXLTKL0020} proposed a series of end-to-end neural network based speech synthesis models and substantially improved the naturalness of synthetic speech. Typically, such models take a text sequence as input and output a Mel-spectrogram, which is subsequently converted to waveform audio using a neural vocoder such as WaveGlow \cite{DBLP:conf/icassp/PrengerVC19} or WaveRNN \cite{DBLP:conf/icml/KalchbrennerESN18}.
\vspace{-0.5cm}
\subsection{Few-shot Speaker Adaptation}
\vspace{-0.2cm}
Although there are many existing speech synthesis methods, they are not directly applicable to our application. Building a speech synthesis system from scratch requires dozens of hours of high-quality recording with corresponding transcriptions \cite{DBLP:journals/corr/abs-2106-15561}, and most existing lecture recordings may not satisfy this requirement. Although some methods such as SV2TTS \cite{DBLP:conf/nips/JiaZWWSRCNPLW18} can clone voice from a small amount of data, their quality remains questionable.

To address this issue, we incorporate few-shot speaker adaptation \cite{DBLP:journals/taslp/YamagishiKNOI09}, a form of transfer learning. It can transfer the basic model with a neutral tone trained on a large-scale multi-speaker dataset to an adapted model with the specific speaker's tone by utilizing a few minutes of audio data. There are two primary approaches to implementing speaker adaptation. The first is to extract the speaker's features using a pre-trained speaker encoding network and then feed them into a neural speech synthesis network \cite{ DBLP:conf/icml/NachmaniPTW18, DBLP:journals/corr/LiMJLZLCKZ17}. The second approach is using a small single-speaker dataset to fine-tune a pre-trained model trained on large multiple-speaker datasets. As demonstrated by existing studies \cite{DBLP:conf/iclr/ChenASBRZWCTLGO19, DBLP:conf/nips/ArikCPPZ18}, the second method is capable of producing higher-quality synthetic speech and our work is based on this approach.

\vspace{-0.5cm}
\subsection{Talking-head Video Generation}
\vspace{-0.2cm}

Given a portrait image or video clip, talking-head video generation refers to the problem of generating a video corresponding to the specific audio. This type of video is frequently used in online education, YouTube video logs and television shows. Existing studies may be classified into two categories: constrained talking-head generation and unconstrained talking-head generation. The constrained talking-head generation methods \cite{DBLP:journals/tog/SuwajanakornSK17, DBLP:journals/corr/abs-1801-01442} are trained on a single speaker datasets, so they are incapable of synthesizing different identities or voices. Additionally, this kind of approach requires a large quantity of training data from a single speaker, which is unsuitable for our application. The unconstrained generation methods \cite{DBLP:journals/tog/0009HSEKL20, DBLP:conf/mm/PrajwalMNJ20} can generate talking-head videos of different speakers using a single model, without training data from any specific speaker. Our approach focuses on unconstrained video generation since this scenario is more suited to our application.
\vspace{-0.4cm}
\section{METHOD}
\label{sec:typestyle}
\vspace{-0.1cm}
\subsection{System pipeline}
\vspace{-0.2cm}

\begin{figure}[t]
\centering
\setlength{\abovecaptionskip}{-0.5cm}
\includegraphics[width=0.45\textwidth]{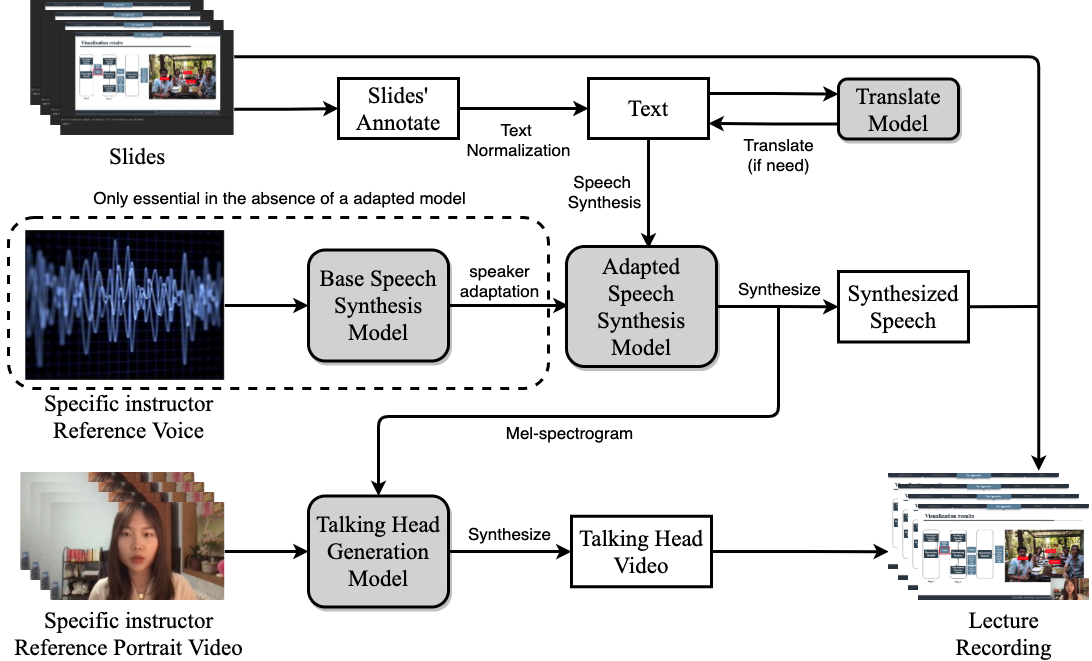}
\caption{System pipeline}
\label{Fig.T2F} 
\vspace{-0.6cm}
\end{figure}

Fig. \ref{Fig.T2F} depicts our end-to-end lecture video generation system. Given annotated slides, we first use the text normalizer to revise abbreviated phrases, numerals, or symbols in the annotation text. This phase can translate written style text to spoken style text, for example, converting ``\$'' to ``dollar.'' If necessary, the text will then be translated to the target language using a translation model. Otherwise, it will be utilized as the direct input to the instructor's adapted speech synthesis model. The instructor's adapted speech synthesis model can generate a cloned voice according to the input text, and the adapted model can be obtained by applying our few-shot speaker adaptation strategy to the base speech synthesis model. The intermediate output of the speech synthesis module, Mel-spectrogram, will be utilized as input to our talking-head generating model, together with the instructor's reference portrait video. This model can generate talking-head videos matching the previously synthesized speech. Finally, lecture videos are obtained by combining the synthesized speech, talking-head video and the slides.

\vspace{-0.4cm}
\subsection{Speech Synthesizer}
\vspace{-0.2cm}

\begin{figure}[t]
\centering
\includegraphics[width=0.45\textwidth]{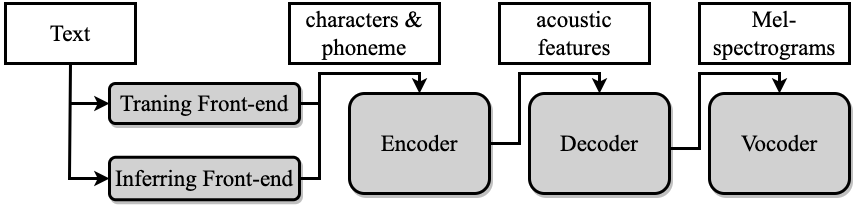}
\caption{Structure of Speech Synthesizer}
\label{Fig.TTS} 
\vspace{-0.4cm}
\end{figure}

\textit{\textbf{Overall Structure:}} As shown in Fig. \ref{Fig.TTS}, our speech synthesizer is based on Tacotron \cite{DBLP:conf/interspeech/WangSSWWJYXCBLA17}. It contains four parts: a dual-channel front-end, an encoder, a decoder and a vocoder.  The dual-channel front-end transforms text sequences to characters and phoneme mixed sequences according to the input type. Following that, the encoder transforms the mixed sequences into acoustic feature sequences, and the decoder generates Mel-spectrograms from the acoustic feature sequences. Finally, the vocoder generates waveform audio files from Mel-spectrogram, and in our implementation, we used the off-the-shelf vocoder: WaveRNN \cite{DBLP:conf/icml/KalchbrennerESN18}.

\textit{\textbf{Randomized Phoneme Replace Training:}} Although the encoder can learn the mapping from words to acoustic features from the dataset, there is no dataset containing all words in the world. Missing words in the dataset will cause mispronunciation \cite{DBLP:conf/interspeech/TaylorR19}. A conventional way is to use a phoneme dictionary to transform words into phonemes before sending them into the encoder. Then the model only needs to learn the mapping from the limited phonemes to acoustic features. However, this will lead to another issue: the encoder cannot deal with neologisms not included in the phoneme dictionary when inferencing. To address this issue, we provide separate front-ends for training and inferencing. During the training phase, each word has a fixed probability of being replaced by its phoneme. The encoder's input will be a character and phoneme mixed sequence, so that the model can learn not only the mapping from phoneme to acoustic feature but also the implicit mapping from character to acoustic feature. Then for the inferencing phase, if the word is in the phoneme dictionary, we will replace it with its phoneme. Otherwise, we keep the characters. This method can ensure the accuracy of most words' pronunciation while ensuring that the encoder can handle neologisms that are not in the phoneme dictionary.

\begin{figure}[t]
\centering 
\setlength{\abovecaptionskip}{0.1cm}
\subfigure[Well-trained ]{
\label{Fig.sub.1}
\includegraphics[width=0.16\textwidth]{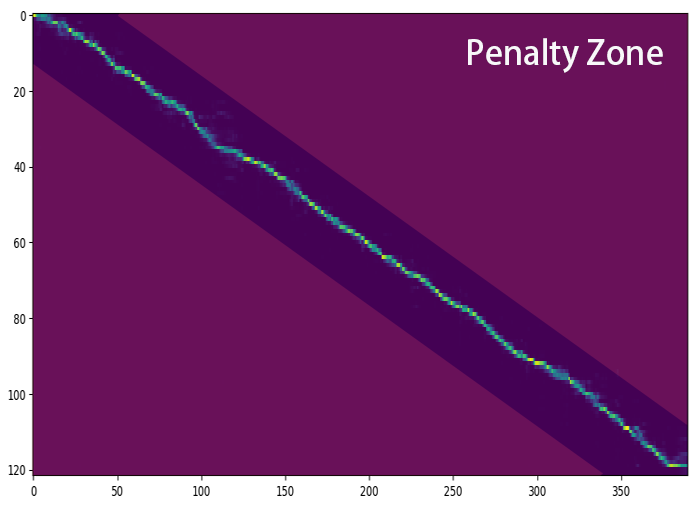}}
\subfigure[Non Well-trained ]{
\label{Fig.sub.2}
\includegraphics[width=0.16\textwidth]{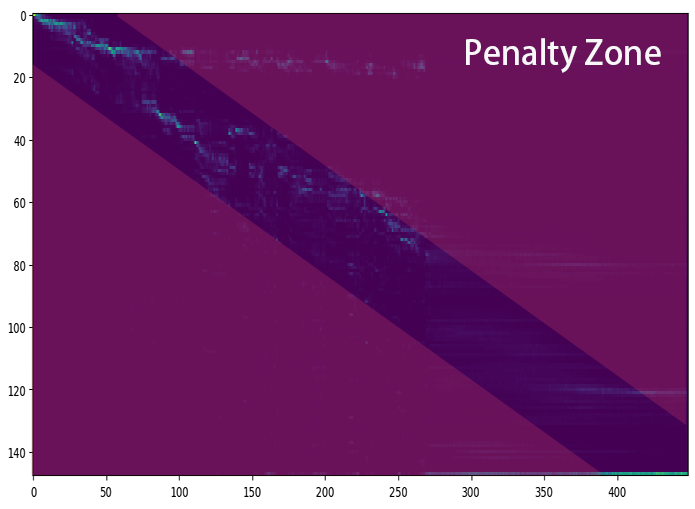}}
\subfigure[Penalty matrix]{
\label{Fig.sub.3}
\includegraphics[width=0.13\textwidth]{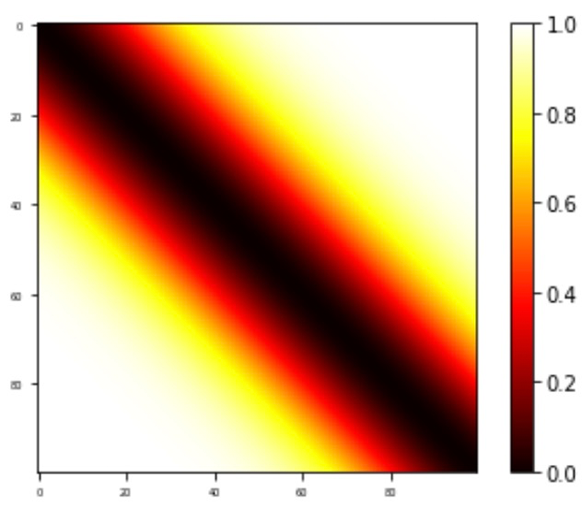}}
\caption{Attention matrix and penalty matrix \(P\)}
\label{Fig.main}
\vspace{-0.6cm}
\end{figure}

\textit{\textbf{Few-shot Speaker Adaptation Strategy:}} Training a speech synthesizer from scratch would require a large amount of high-quality voice data from a specific instructor, which is very difficult to obtain. On the other hand, training on a large-scale multi-speaker dataset allows the pre-trained encoder to generate generic acoustic features for all speakers in the same language. Thus, we only need to adapt the decoder to fine-tune the neutral accent learned from the multi-speaker dataset to a specific speaker's accent using a small training set. Additionally, we discovered that a universal vocoder does not work well with every speaker. It still has a slight effect on the naturalness of the synthesized speech; hence, the vocoder should participate in speaker adaption for better results. To achieve the speaker adaption, we shuffle the multi-speaker dataset and ensure that each training batch has an equal amount of data from various speakers to construct a well-trained model with an average neutral intonation. Following that, we freeze the encoder for the single speaker adaption and directly fine-tune the decoder using the ground truth Mel-spectrogram. Finally, we fine-tune the vocoder following a similar step. This strategy can achieve single speaker adaptation using a very limited training set.

\textit{\textbf{Extra Attention Penalty:}} The attention mechanism is quite important to the decoder. As a well-trained decoder, its attention matrix should visualize as a concentrated diagonal line as shown in Fig. \ref{Fig.sub.1} and Fig. \ref{Fig.sub.2} shows a poorly-trained example. Inspired by  \cite{DBLP:conf/icassp/TachibanaUA18}, in order to speed up the speaker adaptation step, we add an extra penalty of attention matrix in the loss function \(Loss_{att}=  Att\cdot P\), where \(Att\) is the attention matrix and \(P\) is the penalty matrix \(P_{n,t}=1-exp(-k^2\times (n/N-t/T))\) as shown in Fig. \ref{Fig.sub.3} and \(k=3.5\). When the matrix exceeds the penalty zone, it will get a larger loss, and then the back-propagation will try to correct it, making the module converge faster in the speaker adaptation step.

\vspace{-0.3cm}
\subsection{GAN-based Lip Generator}
\vspace{-0.1cm}

\begin{figure}[t]
\setlength{\abovecaptionskip}{-0.2cm}
\centering
\includegraphics[width=0.45\textwidth]{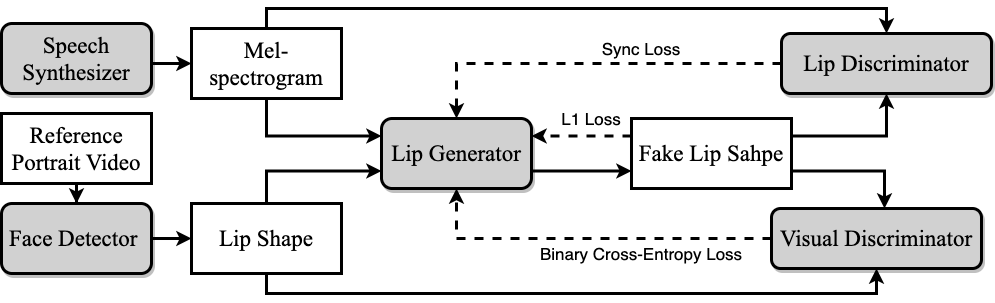}
\caption{Lip Generator structure}
\label{Fig.Lip} 
\vspace{-0.1cm}
\end{figure}

\textit{\textbf{Overall Structure:}} Our lip generator is based on Wav2lip \cite{DBLP:conf/mm/PrajwalMNJ20}. It is a GAN-based model that can directly transform the lip shape of a dynamic video to match the target voice. As shown in Fig. \ref{Fig.Lip}, it contains one generator and two discriminators: the Lip Generator can directly generate lip shape from the Mel-spectrogram; the Lip Discriminator is a separately pre-trained discriminator that is frozen when training the generator, and it can determine whether the Mel-spectrogram of each frame matches the lip shape; and the Visual Discriminator can identify ground truth lip shape and generated lip shape, which is trained with the generator.

\textit{\textbf{Reference Portrait Video Augmentation:}} Because the lecture video is usually much longer than the specific instructor's reference portrait video, if we concatenate the original reference video as loops as Wav2lip does, there will be an obvious gap between the first frame and last frame of the reference portrait video, which will cause unnatural pixel movement between the concatenated frames. Thus we design a video temporal augmentation algorithm, as listed in Algorithm 1. Specifically, we first randomly set a starting point in the range \([0,r \times t]\) and end point in the range \([(1-r) \times t,t]\), where \(t\) is the length of the reference video and \(r\) is an adjustable ratio to constrain the range of values for the starting point and end points. Then crop out the video in the interval, then flip the video and set the previous end point as the new starting point. After that, we randomly select a new end point in the value range \([(1-r) \times t,t]\), crop out the video in the interval again and attach it to the end of the previous one, then repeat this iteratively to generate portrait videos without any gap between video frames. In our experiments, \([0.1,0.3]\) is a suitable value range of \(r\) for the short reference videos and \([0.1,0.4]\) is a suitable value range of \(r\) for the long reference videos. We set \(r=0.2\) in our implementation.

\setlength{\textfloatsep}{5pt}
\begin{algorithm}[t]
\caption{Portrait Video Augmentation}\label{Video Augmentation}
\KwData{reference portrait video frame sequence $P_{t}$ with length $t$, target video length $t'$, adjustable constrain ratio $r$}
\KwResult{augmented portrait video frame sequence $P'_{t'}$}
$I_s\leftarrow rand(0,r \times t)$; $I_e\leftarrow rand((1-r) \times t,t)$;$I=0$\;
\While{$I\leq t'$}{$P'_I=P_{I_s}$;$I_s\leftarrow I_s+1$;$I\leftarrow I+1$\;
\If{$I_s = I_e$}{$P_t=Reverse(P_t)$;
$I_s\leftarrow t - I_s$; $I_e\leftarrow rand((1-r) \times t,t)$\;}
}

\end{algorithm}

\vspace{0.2cm}
\section{EXPERIMENT AND RESULT}
\label{sec:majhead}
\vspace{-0.2cm}
\subsection{Speech Synthesis Evaluation}
\vspace{-0.2cm}
For speech synthesis, we take SV2TTS \cite{DBLP:conf/nips/JiaZWWSRCNPLW18} and Tacotron  \cite{DBLP:conf/interspeech/WangSSWWJYXCBLA17} as our comparison methods. Our base model was developed on LibriSpeech \cite{DBLP:journals/corr/abs-2104-03006} and VCTK \cite{DBLP:conf/ococosda/VeauxYK13}, trained for 120k steps using the Adam optimizer and a step-based learning rate. Then for each unseen speaker, we collected around 40 utterances (around 5 minutes), keeping 20\% of the samples as the test set and the rest as the speaker adaptation set. After training the speaker adaptation for 2000 steps with $lr=3e-5$, then synthesize speeches from the ground truth transcription of the test set. For SV2TTS, we use the whole speaker adaptation set as the speaker encoder's input.

\textit{\textbf{Naturalness.}} We evaluated the naturalness of synthesized speech by mean opinion score (MOS) \cite{DBLP:journals/csl/ViswanathanV05}, which is an evaluation indicator by many text-to-speech works \cite{DBLP:conf/icassp/MossAPGB20, DBLP:journals/corr/abs-2005-05642}. We collected the scores from 40 college students, marked from 1(worst) to 5(best). The results are shown with 95\% confidence interval. 

\textit{\textbf{Speaker Similarity.}} We evaluated the speaker similarity by the average cosine similarity of the 256-dimensional speaker feature between the ground truth and the synthesized speeches, the speaker feature is extracted by Resemblyzer \cite{DBLP:conf/icassp/WanWPL18}.

Table 1 shows the speech synthesis evaluation result. When comparing the speaker similarity with other methods, the speaker similarity of the result produced by our method is 2.3\% higher than SV2TTS, 0.7\%  higher than our method without attention penalty and 40\% higher than Tacotron without speaker adaption. This result shows that our method can produce a better matching voice and the attention penalty can help our model converge better in the same number of steps. As for naturalness, the score of our method is 21.5\% higher than SV2TTS, 3\% higher than Tacotron and 1.7\% higher than our method without attention penalty. This result shows that our method can produce more natural speech and obtain a similar naturalness of none speaker adapted methods.

\vspace{-0.3cm}

\begin{center}
\textbf{Table 1}~~Speech synthesis evaluation results.\\
\setlength{\tabcolsep}{2mm}{
\begin{tabular}{ccccc}
\toprule
Method&Speaker Sim$\uparrow$&Naturalness$\uparrow$&\\
\midrule
Ground truth&1.000&4.37$\pm$0.07&\\
Default&\textbf{0.920}&\textbf{4.00$\pm$0.03}&\\
w/o $Penalty_{att}$&0.914&3.93$\pm$0.04&\\
SV2TTS &0.899&3.29$\pm$0.07&\\
Tacotron &0.656&3.97$\pm$0.05& \\
\bottomrule
\end{tabular}}
\end{center}

\vspace{-0.6cm}
\subsection{Talking-face Generation Evaluation}
\vspace{-0.2cm}

We compare our talking-face generation method with MakeIt-Talk \cite{DBLP:journals/tog/0009HSEKL20} and PC-AVS \cite{DBLP:conf/cvpr/ZhouSWL0021}. Our model is trained on the sync-corrected and cleaned VoxCeleb2 Dataset \cite{DBLP:conf/interspeech/ChungNZ18}. For each experiment, we generate 4 talking-head videos by 4 ground truth speeches audio and a reference portrait video from an unseen person. The duration of the reference portrait video is 5 seconds and all the speeches are around 10 to 20 seconds. For MakeItTalk and PC-AVS, we used their official implementations and pre-trained models.

\textit{\textbf{Lip Sync Confidence.}} The lip sync confidence is measured by Syncnet \cite{DBLP:conf/accv/ChungZ16a}, which measures the lip-sync confidence between the generated frames and the ground truth audio segment. 

\textit{\textbf{Naturalness.}} The naturalness is measured by the mean opinion score. We collected the scores from 40 college students, marked from 1(worst) to 5(best). The results are shown with 95\% confidence interval.

\vspace{-0.3cm}

\begin{center}
\textbf{Table 2}~~Lip generation evaluation results.\\
\setlength{\tabcolsep}{2mm}{
\begin{tabular}{ccc}
\toprule
Method&Lip Conf$\uparrow$&Nat$\uparrow$\\
\midrule
Default&8.531&\textbf{4.36$\pm$0.10}\\
w/o Augmentation&\textbf{8.537}&4.15$\pm$0.08\\
WAV2LIP &8.524&4.11$\pm$0.09\\
MakeItTalk &4.594&3.32$\pm$0.10\\
PC-AVS &3.782&3.12$\pm$0.11\\
\bottomrule
\end{tabular}}
\end{center}

The results of lip generation are shown in Table 2. Benefitting from wav2lip's structure, our method achieved nearly double lip-sync confidence than MakeItTalk and PC-AVS. At the same time, our video augmentation method improved the naturalness of the generated video, 5\%, 6\%, 31\%, 39\% comparing our method without augmentation, WAV2LIP, MakeItTalk and PC-AVS, respectively. We also notice that excessively vigorous motions in reference videos may cause unsatisfactory results.

\vspace{-0.4cm}
\section{CONCLUSION}
\label{sec:foot}
\vspace{-0.2cm}
We proposed an end-to-end lecture video generation system that can produce realistic lecture videos automatically from annotated slides. A voice synthesizer with a few-shot speaker adaptation strategy and a GAN-based talking-head generator are the core components of our system. We evaluated various metrics of the synthesized speech and talking-head videos, and the evaluation reveals that our approach outperforms existing methods on all metrics.

\vfill\pagebreak

\bibliographystyle{IEEEbib}
\bibliography{strings,refs}

\begin{thebibliography}{00}
\footnotesize
\addtolength{\itemsep}{-0.5 em}

\bibitem{DBLP:conf/interspeech/WangSSWWJYXCBLA17}

Y.~Wang, R.~J. Skerry{-}Ryan, D.~Stanton, Y.~Wu, R.~J. Weiss, N.~Jaitly,
  Z.~Yang, Y.~Xiao, Z.~Chen, S.~Bengio, Q.~V. Le, Y.~Agiomyrgiannakis,
  R.~Clark, and R.~A. Saurous, ``Tacotron: Towards end-to-end speech
  synthesis,'' in \emph{Proc. Interspeech},
  2017, pp. 4006--4010.


\bibitem{DBLP:journals/corr/abs-2106-15561}

X.~Tan, T.~Qin, F.~K. Soong, and T.~Liu, ``A survey on neural speech
  synthesis,'' \emph{arXiv preprint}, 2021, \emph{arXiv:2106.15561}.

\bibitem{DBLP:conf/icassp/ShenPWSJYCZWRSA18}

J.~Shen, R.~Pang, R.~J. Weiss, M.~Schuster, N.~Jaitly, Z.~Yang, Z.~Chen,
  Y.~Zhang, Y.~Wang, R.~Ryan, R.~A. Saurous, Y.~Agiomyrgiannakis, and Y.~Wu,
  ``Natural {TTS} synthesis by conditioning wavenet on {MEL} spectrogram
  predictions,'' in \emph{Proc. ICASSP}, 2018, pp.
  4779--4783.


\bibitem{DBLP:conf/icml/WangSZRBSXJRS18}

Y.~Wang, D.~Stanton, Y.~Zhang, R.~J. Skerry{-}Ryan, E.~Battenberg, J.~Shor,
  Y.~Xiao, Y.~Jia, F.~Ren, and R.~A. Saurous, ``Style tokens: Unsupervised
  style modeling, control and transfer in end-to-end speech synthesis,'' in
  \emph{Proc. ICML}, 2018, pp. 5167--5176.


\bibitem{DBLP:conf/interspeech/YuLH0WXLTKL0020}

C.~Yu, H.~Lu, N.~Hu, M.~Yu, C.~Weng, K.~Xu, P.~Liu, D.~Tuo, S.~Kang, G.~Lei,
  D.~Su, and D.~Yu, ``Durian: Duration informed attention network for speech
  synthesis,'' in \emph{Proc. Interspeech}, 2020, pp. 2027--2031.


\bibitem{DBLP:conf/icassp/PrengerVC19}

R.~Prenger, R.~Valle, and B.~Catanzaro, ``Waveglow: {A} flow-based generative
  network for speech synthesis,'' in \emph{Proc. ICASSP},
  2019, pp. 3617--3621.


\bibitem{DBLP:conf/icml/KalchbrennerESN18}

N.~Kalchbrenner, E.~Elsen, K.~Simonyan, S.~Noury, N.~Casagrande, E.~Lockhart,
  F.~Stimberg, A.~van~den Oord, S.~Dieleman, and K.~Kavukcuoglu, ``Efficient
  neural audio synthesis,'' in \emph{Proc. ICML}, 2018, pp. 2415--2424.


\bibitem{DBLP:conf/icml/NachmaniPTW18}

E.~Nachmani, A.~Polyak, Y.~Taigman, and L.~Wolf, ``Fitting new speakers based
  on a short untranscribed sample,'' in \emph{Proc. ICML}, 2018, pp. 3680--3688.


\bibitem{DBLP:journals/corr/LiMJLZLCKZ17}

C.~Li, X.~Ma, B.~Jiang, X.~Li, X.~Zhang, X.~Liu, Y.~Cao, A.~Kannan, and Z.~Zhu,
  ``Deep speaker: an end-to-end neural speaker embedding system,'' \emph{arXiv preprint}, 2017, \emph{arXiv:1705.02304}.


\bibitem{DBLP:journals/taslp/YamagishiKNOI09}

J.~Yamagishi, T.~Kobayashi, Y.~Nakano, K.~Ogata, and J.~Isogai, ``Analysis of
  speaker adaptation algorithms for hmm-based speech synthesis and a
  constrained {SMAPLR} adaptation algorithm,'' \emph{{IEEE} Trans. Speech Audio
  Process.}, vol.~17, no.~1, 2009, pp. 66--83.


\bibitem{DBLP:conf/nips/ArikCPPZ18}

S.~{\"{O}}. Arik, J.~Chen, K.~Peng, W.~Ping, and Y.~Zhou, ``Neural voice
  cloning with a few samples,'' in \emph{Proc. NeurIPS}, 2018, pp. 10\,040--10\,050.

\bibitem{DBLP:conf/icassp/ChungWHZS19}

Y.~Chung, Y.~Wang, W.~Hsu, Y.~Zhang, and R.~J. Skerry{-}Ryan, ``Semi-supervised
  training for improving data efficiency in end-to-end speech synthesis,'' in
  \emph{Proc. ICASSP}, 2019, pp. 6940--6944.


\bibitem{calvo2020educating}
S.~Calvo, F.~Lyon, A.~Morales, and J.~Wade, ``Educating at scale for
  sustainable development and social enterprise growth: The impact of online
  learning and a massive open online course (mooc),'' \emph{Sustainability},
  vol.~12, no.~8, p. 3247, 2020.

\bibitem{matsumoto2019ai}
K.~Matsumoto, T.~Nakanishi, T.~Sakawa, K.~Onodera, S.~Orimo, and H.~Kobayashi,
  ``Ai-josyu: Thinking support system in class by real-time speech recognition
  and keyword extraction,'' \emph{EMITTER International Journal of Engineering
  Technology}, 2019, vol.~7, no.~1, pp. 366--383.

\bibitem{DBLP:journals/csl/ViswanathanV05}

M.~Viswanathan and M.~Viswanathan, ``Measuring speech quality for
  text-to-speech systems: development and assessment of a modified mean opinion
  score {(MOS)} scale,'' \emph{Comput. Speech Lang.}, 2005, vol.~19, no.~1, pp.
  55--83.


\bibitem{DBLP:conf/icassp/MossAPGB20}

H.~B. Moss, V.~Aggarwal, N.~Prateek, J.~Gonz{\'{a}}lez, and R.~Barra{-}Chicote,
  ``{BOFFIN} {TTS:} few-shot speaker adaptation by bayesian optimization,'' in
  \emph{Proc. ICASSP}, 2020, pp. 7639--7643.

\bibitem{DBLP:journals/corr/abs-2005-05642}

Z.~Zhang, Q.~Tian, H.~Lu, L.~Chen, and S.~Liu, ``Adadurian: Few-shot adaptation
  for neural text-to-speech with durian,'' \emph{arXiv preprint},
  2020, \emph{arXiv:2005.05642}.


\bibitem{DBLP:journals/corr/abs-2104-03006}

A.~Zeyer, A.~Merboldt, W.~Michel, R.~Schl{\"{u}}ter, and H.~Ney, ``Librispeech
  transducer model with internal language model prior correction,'' in \emph{Proc. Interspeech}, 2021, pp. 2052--2056.


\bibitem{DBLP:conf/ococosda/VeauxYK13}

C.~Veaux, J.~Yamagishi, and S.~King, ``The voice bank corpus: Design,
  collection and data analysis of a large regional accent speech database,'' in
  \emph{Proc. COCOSDA}, 2013, pp. 1--4.


\bibitem{DBLP:conf/interspeech/ChungNZ18}

J.~S. Chung, A.~Nagrani, and A.~Zisserman, ``Voxceleb2: Deep speaker
  recognition,'' in \emph{Proc. Interspeech}, 2018, pp. 1086--1090.


\bibitem{DBLP:conf/accv/ChungZ16a}

J.~S. Chung and A.~Zisserman, ``Out of time: Automated lip sync in the wild,''
  in \emph{Proc. ACCV Workshops}, 2016, pp. 251--263.


\bibitem{DBLP:conf/icassp/WanWPL18}

L.~Wan, Q.~Wang, A.~Papir, and I.~Lopez{-}Moreno, ``Generalized end-to-end loss
  for speaker verification,'' in \emph{Proc. ICASSP}, 2018, pp.
  4879--4883.


\bibitem{DBLP:journals/tog/SuwajanakornSK17}

S.~Suwajanakorn, S.~M. Seitz, and I.~Kemelmacher{-}Shlizerman, ``Synthesizing
  obama: learning lip sync from audio,'' \emph{{ACM} Trans. Graph.}, 2017, vol.~36,
  no.~4, pp. 95:1--95:13.


\bibitem{DBLP:journals/corr/abs-1801-01442}

R.~Kumar, J.~Sotelo, K.~Kumar, A.~de~Br{\'{e}}bisson, and Y.~Bengio,
  ``Obamanet: Photo-realistic lip-sync from text,'' in
  \emph{Proc. NeurIPS Workshops}, 2017, pp. 5998--6008.


\bibitem{DBLP:conf/mm/PrajwalMNJ20}

K.~R. Prajwal, R.~Mukhopadhyay, V.~P. Namboodiri, and C.~V. Jawahar, ``A lip
  sync expert is all you need for speech to lip generation in the wild,'' in
  \emph{Proc. ACMMM}, 2020, pp. 484--492.


\bibitem{DBLP:conf/icassp/TachibanaUA18}

H.~Tachibana, K.~Uenoyama, and S.~Aihara, ``Efficiently trainable
  text-to-speech system based on deep convolutional networks with guided
  attention,'' in \emph{Proc. ICASSP}, 2018, pp.
  4784--4788.


\bibitem{DBLP:journals/tog/0009HSEKL20}

Y.~Zhou, X.~Han, E.~Shechtman, J.~Echevarria, E.~Kalogerakis, and D.~Li,
  ``Makelttalk: speaker-aware talking-head animation,'' \emph{{ACM} Trans.
  Graph.}, 2020, vol.~39, no.~6, pp. 221:1--221:15.


\bibitem{DBLP:conf/nips/VaswaniSPUJGKP17}

A.~Vaswani, N.~Shazeer, N.~Parmar, J.~Uszkoreit, L.~Jones, A.~N. Gomez,
  L.~Kaiser, and I.~Polosukhin, ``Attention is all you need,'' in
  \emph{Proc. NeurIPS}, 2017, pp. 5998--6008.


\bibitem{DBLP:conf/icmcs/XuWLLZSH19}

C.~Xu, R.~Wang, S.~Lin, X.~Luo, B.~Zhao, L.~Shao, and M.~Hu, ``Lecture2note:
  Automatic generation of lecture notes from slide-based educational videos,''
  in \emph{Proc. ICME}, 2019, pp. 898--903.


\bibitem{lederman2020teaching}
D.~Lederman, ``How teaching changed in the (forced) shift to remote learning,''
  \emph{Inside Higher Ed}, 2020, vol.~22.

\bibitem{DBLP:conf/interspeech/TaylorR19}

J.~Taylor and K.~Richmond, ``Analysis of pronunciation learning in end-to-end
  speech synthesis,'' in \emph{Proc. Interspeech}, 2019, pp. 2070--2074.

\bibitem{DBLP:conf/cvpr/ZhouSWL0021}

H.~Zhou, Y.~Sun, W.~Wu, C.~C. Loy, X.~Wang, and Z.~Liu, ``Pose-controllable
  talking face generation by implicitly modularized audio-visual
  representation,'' in \emph{Proc. CVPR}, 2021, pp. 4176--4186.

\bibitem{DBLP:conf/nips/JiaZWWSRCNPLW18}
Y.~Jia, Y.~Zhang, R.~J. Weiss, Q.~Wang, J.~Shen, F.~Ren, Z.~Chen, P.~Nguyen,
  R.~Pang, I.~Lopez{-}Moreno, and Y.~Wu, ``Transfer learning from speaker
  verification to multispeaker text-to-speech synthesis,'' in \emph{Proc. NeurIPS},
  2018, pp. 4485--4495.

\bibitem{DBLP:conf/iclr/ChenASBRZWCTLGO19}
Y.~Chen, Y.~M. Assael, B.~Shillingford, D.~Budden, S.~E. Reed, H.~Zen, Q.~Wang,
  L.~C. Cobo, A.~Trask, B.~Laurie, {\c{C}}.~G{\"{u}}l{\c{c}}ehre, A.~van~den
  Oord, O.~Vinyals, and N.~de~Freitas, ``Sample efficient adaptive
  text-to-speech,'' in \emph{Proc. ICLR}, 2019.
  
\end{thebibliography}

\end{document}